\documentclass[12pt]{article}
\usepackage{graphicx}
\usepackage{amssymb}
\usepackage{epstopdf}
\title{Stokes Theorem and the Equations of GRMHD}
\author{ Jean-Pierre De Villiers\\
\footnotesize{email: jpd\underline{ }asph@mac.com}}

\begin{document}
\maketitle
\normalsize

\abstract{
This paper provides an alternate derivation of the equations used in the 
GRMHD code of De Villiers \& Hawley using Stokes Theorem. This
derivation places the published form of the discretized equations of
GRMHD in a broader context, and suggests possible future directions for 
numerical work.}
\section{Introduction}
\parskip=12pt

The general relativistic magnetohydrodynamic (GRMHD) 
code of De Villiers \& Hawley (2003; DH) has been used to study the
accretion of matter onto black holes. It is a highly stable $3+1$D 
code that
has been used to simulate in great detail the evolution of accretion disks in
the spacetime of Kerr black holes. This code has established the ubiquity of
energetic axial jets in such environments (De Villiers {\it et al.}, 2003, 2005), and has shed light on the magnetic environment produced accretion disks (Hirose {\it et al.}, 2004; De Villiers, 2006).

The refinement of the code through testing and development is
ongoing. One area of study that may help spur future improvements is the
mathematical foundation of the GRMHD equations. There are many threads in
the literature documenting the differential version of the equations. Misner,
Thorne, \& Wheeler (MTW; 1973) provide a clear derivation of the equations of
hydrodynamics. Anile (1989) discusses the extension to magnetohydrodynamics (MHD) in
a special relativistic background. Hawley, Smarr, \& Wilson (1984; HSW) treat 
hydrodynamics in a fixed background (black hole spacetime), while DH
completes the derivation of the complete set of GRMHD equations. 

This paper presents a derivation of the GRMHD equations from the
generalized Stokes Theorem (MTW; see also Page \& Thorne, 1974), and
was motivated by a derivation due to Clarke 
(1996) of the equations of non-relativistic MHD as discretized for the ZEUS code. Clarke provides an integral formulation of the equations of MHD using Stokes Theorem, and clearly
shows the connection between the apparently simplistic finite-differencing
techniques used in ZEUS (and also in the GRMHD code) to the notion of flux 
conservation embodied in Stokes Theorem.

\section{Conservation Laws of GRMHD}

The state of an ideal magnetized fluid on a curved background is described by
a set of conservation laws. The equations
in differential form are
\begin{eqnarray}
\nabla_\mu \,(\rho\,U^\mu) = 0 & & {\rm (continuity)},\\
\nabla_\mu \,T^{\mu \nu} = 0 & & {\rm (energy-momentum)},\\
\nabla_\mu \,{}^{*}F^{\mu \nu} = 0 & & {\rm (induction)},
\end{eqnarray}
where $\nabla_\mu$ is the covariant derivative, $\rho$ is the density, $U^\mu$ the 4-velocity, 
$T^{\mu \nu}$ the energy-momentum tensor,
$F^{\mu \nu}$ the electromagnetic field strength tensor, and its dual is
\begin{eqnarray}
\label{dual.2}
{}^*F^{\mu\,\nu} & = & {1 \over 2}\,\epsilon^{\mu\,\nu\,\delta\,\gamma} 
 F_{\delta \gamma} ,
\end{eqnarray}
where $\epsilon^{\mu\,\nu\,\delta\,\gamma}
= -(1/\sqrt{-g})\,[\mu\,\nu\,\delta\,\gamma]$ is the contravariant
form of the Levi-Civita tensor. The energy-momentum tensor is
\begin{equation} \label{tmndef}
{T}^{\mu\,\nu} = \rho\,h\,{U}^{\mu}\,{U}^{\nu}+
 P\,{g}^{\mu\,\nu} + {1 \over 4\,\pi} \left({F^\mu}_\alpha\,F^{\nu \alpha} 
 - {1 \over 4} F_{\alpha \beta}\,F^{\alpha \beta}\,g^{\mu \nu} \right) .
\end{equation}
where $h=1 + \epsilon + P/\rho$ is the specific 
enthalpy, with $\epsilon$ the specific internal energy and 
$P=\rho\,\epsilon\,(\Gamma-1)$ the ideal gas pressure ($\Gamma$ is the
adiabatic exponent). 

Following Lichnerowicz (1967) and Anile (1989), define the magnetic 
induction and the electric field by projecting onto the four-velocity,
\begin{eqnarray}
\label{bdef.2}
B^{\mu} & = & {}^*F^{\mu\,\nu}\,U_\nu ,\\
\label{edef.2}
E^{\mu} & = & F^{\mu \nu}\,U_\nu .
\end{eqnarray}
The flux-freezing condition, wherein the electric field in the fluid rest
frame is zero, implies $F^{\mu \nu}\,U_\nu = 0$.
Combine the definition of the magnetic induction (\ref{bdef.2}) with 
(\ref{dual.2}) and flux freezing to obtain
\begin{equation}
\label{fmunudef.2}
F_{\mu\,\nu} = \epsilon_{\alpha\,\beta\,\mu\,\nu}\,B^{\alpha}\,U^\beta ,
\end{equation}
where 
$\epsilon_{\mu\,\nu\,\delta\,\gamma}=\sqrt{-g}\,[\mu\,\nu\,\delta\,\gamma]$.
The orthogonality 
condition 
\begin{equation}\label{bnorm}
B^{\mu}\,U_{\mu} = 0
\end{equation}
follows directly from (\ref{bdef.2}) and the anti-symmetry of $F^{\mu \nu}$.

Using these results, it is possible to rewrite the 
electromagnetic portion of the energy-momentum tensor to obtain 
\begin{equation}
T^{\mu \nu} =  \left(\rho\,h+{\|b\|}^2\right)\,U^\mu\,U^\nu + 
\left(P+{{\|b\|}^2 \over 2}\right)\,g^{\mu \nu} - b^\mu\,b^\nu
\end{equation}
where ${\|b\|}^2=b^\mu\,b_\mu$ is the magnetic field 
intensity, and $b^\mu = B^\mu/(4\,\pi)$ is the
magnetic field four-vector.

In addition, we make the identification
\begin{equation}\label{ctvars}
{\cal{B}}^r      = F_{\phi \theta} \, , \,
{\cal{B}}^\theta = F_{r \phi} \, , \,
{\cal{B}}^\phi   = F_{\theta r} ,
\end{equation}
where ${\cal{B}}^j$ are the Constrained Transport (CT; Evans \& Hawley, 1988)
magnetic field variables. Also,
\begin{eqnarray}\label{maxwell.2c}
F_{t r}      & = & V^\phi\,{\cal{B}}^{\theta} -V^\theta\,{\cal{B}}^{\phi} 
               = {\cal E}^r,\\
\nonumber F_{t \theta} & = & V^r\,{\cal{B}}^{\phi} -V^\phi\,{\cal{B}}^{r}
               = {\cal E}^\theta,\\
\nonumber F_{t \phi}   & = & V^\theta\,{\cal{B}}^{r} - V^r\,{\cal{B}}^{\theta}
               = {\cal E}^\phi,
\end{eqnarray}
where ${\cal E}^i$ are the CT EMFs.

\section{Stokes Theorem}

The generalized form of Stokes Theorem (MTW) relates the divergence of a vector
field (${\cal A}^\mu$) in a 4-volume (${\cal V}$) to the flux of this vector 
field through the oriented surface ($\partial{\cal V}$) bounding this volume:
\begin{equation}\label{stokes}
\int_{\cal V} \nabla_\mu\,{\cal A}^\mu\,d^4\Omega= 
\int_{\partial {\cal V}} {\cal A}^\mu\,d^3\Sigma_\mu 
\end{equation}
where $d^4\Omega$ is the volume element of ${\cal V}$ and $d^3\Sigma_\mu$ is 
the four-vector outward normal to the surface $\partial{\cal V}$.
Since the GRMHD code uses the Boyer-Lindquist coordinates for the Kerr
metric, the time and space coordinates $(t,r,\theta,\phi)$ will be used from
hereon.
In Boyer-Lindquist coordinates, the volume element is
$d^4\Omega = \sqrt{-g}\,dt\,dr\,d\theta\,d\phi$
where $\sqrt{-g} = \alpha\,\sqrt{\gamma}$, 
$\alpha = {(-g^{tt})}^{-1/2}$ is the lapse function,
and $\sqrt{\gamma}$ is the determinant of the spatial 3-metric.

\begin{figure}[htbp]
\begin{center}
\includegraphics[width=4.5in]{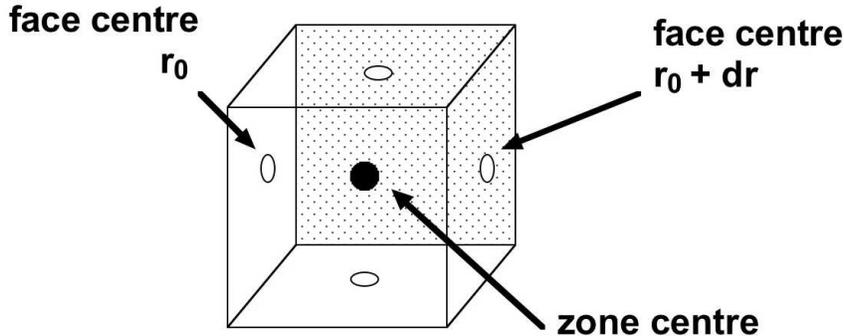}
\caption{{\small Sketch of hypersurfaces bounding the four-dimensional
volume ${\cal V}$.}}
\label{zone}
\end{center}
\end{figure}

Expression (\ref{stokes}) is completely general; in anticipation
of its application to the GRMHD code, take the four-volume ${\cal V}$ 
to be an infinitessimal region whose geometry matches that of a discrete 
$3+1D$ zone in the computational grid. This four-volume has eight well 
defined boundary planes, corresponding to hypersurfaces where one the the 
Boyer-Linquist coordinates is held constant. Figure \ref{zone} presents a 
three-dimensional sketch of the bounding faces, and only the bounding faces in 
the radial direction are labelled. The surface denoted $r_0$ represents the 
hypersurface of constant $r=r_0$, the ``lower''
bounding hypersurface of ${\cal V}$ in the radial direction. The surface
denoted $r_0 + dr$ represents the hypersurface of constant $r=r_0+dr$, the 
``upper'' bounding hypersurface of ${\cal V}$ in the radial direction.
Similarly, the hypersurfaces of constant polar angle are denoted 
$\theta_0$ and $\theta_0 + d\theta$; and hypersurfaces of constant azimuthal angle, $\phi_0$ and $\phi_0 + d\phi$. The generalized form of Stokes Theorem
also deals with the time coordinate on an equal footing, so there are also surfaces of constant
time, $t_0$ and $t_0 + dt$, bounding ${\cal V}$.

%%
%% Figure
%%
%%
%% END Figure
%%

Figure \ref{zone} also introduces some additional terminology for the
staggered mesh used in the GRMHD code: scalar quantities
(e.g. $\rho$) reside on zone centres, whereas vector
quantities (e.g. ${\cal B}^i$) reside on boundary surfaces (zone faces). In addition, the discrete grid associates the face-centred quantities to the ``lower'' bounding hypersurface (e.g. ${\cal B}^r$ associated with the zone of Figure \ref{zone} lies on the centre of the $r_0$ hypersurface, while the ${\cal B}^r$ on the 
$r_0+dr$ hypersurface is associated with the next zone in the radial direction). 

\section{Discretizing the Integral Equations}

The discretization of the integral form of the equations of GRMHD can
be illustrated by considering a
conservation law for some arbitrary vector field,
$\nabla_\mu {\cal A}^\mu =0$, which can be expressed using Stokes
Theorem (\ref{stokes}) as an integral over ${\partial {\cal V}}$:
\begin{equation} 
\int_{\partial {\cal V}} {\cal A}^\mu\,d^3\Sigma_\mu = 0.  
\end{equation}
Using the geometry of volume ${\cal V}$ introduced in the previous
section, the conservation law reads,
\begin{eqnarray} \label{flux}
0=\int_{\partial {\cal V}} {\cal A}^\mu\,d^3\Sigma_\mu 
 & = & \int_{r_0+dr} {\cal A}^r\,\sqrt{-g}\,dt\,d\theta\,d\phi -
       \int_{r_0} {\cal A}^r\,\sqrt{-g}\,dt\,d\theta\,d\phi \\
\nonumber
 & + & \int_{\theta_0+d\theta} {\cal A}^\theta\,\sqrt{-g}\,dt\,dr\,d\phi -
       \int_{\theta_0} {\cal A}^\theta\,\sqrt{-g}\,dt\,dr\,d\phi \\
\nonumber
 & + & \int_{\phi_0+d\phi} {\cal A}^\phi\,\sqrt{-g}\,dt\,dr\,d\theta -
       \int_{\phi_0} {\cal A}^\phi\,\sqrt{-g}\,dt\,dr\,d\theta \\
\nonumber
 & + & \int_{t_0+dt} {\cal A}^t\,\sqrt{-g}\,dr\,d\theta\,d\phi  -
       \int_{t_0} {\cal A}^t\,\sqrt{-g}\,dr\,d\theta\,d\phi
\end{eqnarray}
where the subscript on each integral indicates the specific hypersurface (see
Figure \ref{zone}), and each integral symbol in understood to represent a definite triple integral over each bounding hypersurface.

The presence of the time coordinate in the integrals over surfaces of constant 
spatial coordinate serves as a pointed reminder that proper 
centering of physical quantities must be done not only in space but in time as 
well. If for no other reason, the natural emergence of this important connection between all four coordinates validates the use of
the integral formulation as the starting point for numerical discretization.

In going to a discrete grid, letting $d x^\mu \rightarrow \Delta x^\mu$, 
(\ref{flux}) becomes an approximate relation, and there are many possible
ways to proceed with the discretization. To see how this comes about, consider 
the flux of ${\cal A}^\mu$ through the surface of constant $r_0$,
\begin{equation} \label{flux.2}
\int_{r_0} {\cal A}^r\,\sqrt{-g}\,dt\,d\theta\,d\phi \equiv
\left\{\int_{\phi_0}^{\phi_0+\Delta \phi} \int_{\theta_0}^{\theta _0+\Delta \theta} \int_{t_0}^{t_0+\Delta t} {\cal A}^r\,\sqrt{-g}\,d t\,d \theta\,d \phi \right\}_{r_0} 
\end{equation}
In subsequent expressions, the braces $\left\{\quad\right\}$ will denote
a given hypersurface; this notation
is similar to that used by Page \& Thorne (1974).
Neglecting the integral over time for a moment, the simplest approximation
to (\ref{flux.2}) is the product of the mean radial component of ${\cal A}$
on the surface multiplied by the area of the surface,
\begin{equation} \label{flux.3}
\left\{\int_{\phi_0}^{\phi_0+\Delta \phi} \int_{\theta_0}^{\theta _0+\Delta \theta} {\cal A}^r\,\sqrt{-g}\,d \theta\,d \phi \right\}_{r_0} \approx 
\langle{\cal A}^{\,r}\rangle_{r_0} \left\{\int_{\phi_0}^{\phi_0+\Delta \phi} \int_{\theta_0}^{\theta _0+\Delta \theta} \sqrt{-g}\,d \theta\,d \phi \right\}_{r_0},
\end{equation}
where $\langle{\cal A}^{\,r}\rangle_{r_0}$ represents the mean value of ${\cal A}^r$ on the
hypersurface $r_0$. On a discrete grid, the flux could be approximated
by taking quantities on the zone face as representative of this mean so that
\begin{equation} \label{flux.4}
\left\{\int_{\phi_0}^{\phi_0+\Delta \phi} \int_{\theta_0}^{\theta _0+\Delta \theta} {\cal A}^r\,\sqrt{-g}\,d \theta\,d \phi \right\}_{r_0} \approx 
\left\{ {\cal A}^r \sqrt{-g}\,\Delta \theta\,\Delta \phi\right\}_{r_0},
\end{equation}
where ${\cal A}^r$ and $\sqrt{-g}$ on the right-hand side are understood to be 
evaluated at the centre of the $r_0$ zone face. 

However, equation (\ref{flux.2}) also 
includes the integral over time which must be dealt with in a consistent
manner. In a numerical implementation, the solution at time level $t_0$ 
is known and the solution at time level $t_0+\Delta t$ is what is sought, so
(\ref{flux.2}) cannot be evaluated exclusively with data at the 
known time level $t_0$ without introducing potentially serious numerical errors. 
For consistency, (\ref{flux.2}) should be evaluated in such a way as to involve 
the unknown time level. 
For example, a complete discretization of (\ref{flux.2}) could be written as
\begin{equation} \label{flux.5}
\left\{\int_{\phi_0}^{\phi_0+\Delta \phi} \int_{\theta_0}^{\theta _0+\Delta \theta} \int_{t_0}^{t_0+\Delta t} {\cal A}^r\,\sqrt{-g}\,d t\,d \theta\,d \phi \right\}_{r_0} \approx {1 \over 2}
\left\{ \left[{{\cal A}^r}^{\,(n)}+{{\cal A}^r}^{\,(n+1)} \right]\sqrt{-g}\,\Delta t\,\Delta \theta\,\Delta \phi\right\}_{r_0},
\end{equation}
where solutions at different time levels are distinguished by introducing superscripts; ${{\cal A}^r}^{\,(n)}$ denotes the value of ${\cal A}^r$ at the 
current time $t_0$, and ${{\cal A}^r}^{\,(n+1)}$ at the next time level, 
$t_0 + \Delta t$.

Similar expressions can be derived for the other hypersurfaces of constant
spatial coordinate. This leaves the two hypersurfaces
of constant time. Take the surface of constant $t_0$, the flux through this
surface is
\begin{equation} \label{flux.6}
\int_{t_0} {\cal A}^t\,\sqrt{-g}\,dr\,d\theta\,d\phi \equiv
\left\{\int_{\phi_0}^{\phi_0+\Delta \phi} \int_{\theta_0}^{\theta _0+\Delta \theta} \int_{r_0}^{r_0+\Delta r} {\cal A}^t\,\sqrt{-g}\,d r\,d \theta\,d \phi \right\}_{t_0}, 
\end{equation}
which can be approximated as
\begin{equation} \label{flux.6}
\left\{\int_{\phi_0}^{\phi_0+\Delta \phi} 
       \int_{\theta_0}^{\theta _0+\Delta \theta} 
       \int_{r_0}^{r_0+\Delta r} 
       {\cal A}^t\,\sqrt{-g}\,d r\,d \theta\,d \phi \right\}_{t_0} \approx
 {{\cal A}^t}^{\,(n)}\sqrt{-g}\,\Delta r\,\Delta \theta\,\Delta \phi. 
\end{equation}
Note that for the hypersurfaces of constant time, the brace notation is not used, and the time-components of vectors are taken to be zone-centred quantities.

Combining all these terms, and, in keeping with the GRMHD code, taking 
$\Delta t$ to be constant over the entire grid, at all times, (\ref{flux}) 
can be approximated as
\begin{eqnarray} \label{flux.7}
0 & \approx &
  \,{{{\cal A}^t}^{\,(n+1)} - {{\cal A}^t}^{\,(n)} \over \Delta t}+ {1 \over 2\,\sqrt{-g}} \\
\nonumber
  & \times &  \left(
 {\left\{\sqrt{-g}\,
         \left[{{\cal A}^r}^{\,(n+1)} + {{\cal A}^r}^{\,(n)}\right]
  \right\}_{r_0+\Delta r} -
\left\{\sqrt{-g}\,
         \left[{{\cal A}^r}^{\,(n+1)} + {{\cal A}^r}^{\,(n)}\right]
  \right\}_{r_0} 
  \over \Delta r}\right. \\
\nonumber
  & + &  
 {\left\{\sqrt{-g}\,
         \left[{{\cal A}^\theta}^{\,(n+1)} + {{\cal A}^\theta}^{\,(n)}\right]
  \right\}_{\theta_0+\Delta \theta} -
\left\{\sqrt{-g}\,
         \left[{{\cal A}^\theta}^{\,(n+1)} + {{\cal A}^\theta}^{\,(n)}\right]
  \right\}_{\theta_0} 
  \over \Delta \theta}\\
  & + &  \nonumber \left.
 {\left\{\sqrt{-g}\,
         \left[{{\cal A}^\phi}^{\,(n+1)} + {{\cal A}^\phi}^{\,(n)}\right]
  \right\}_{\phi_0+\Delta \phi} -
\left\{\sqrt{-g}\,
         \left[{{\cal A}^\phi}^{\,(n+1)} + {{\cal A}^\phi}^{\,(n)}\right]
  \right\}_{\phi_0} 
  \over \Delta \phi}  \right).
\end{eqnarray}

Of course, this is but one possible discretization of (\ref{flux}). The
advantage of the integral formulation is that it facilitates writing
flux-conserving numerical routines which achieve proper centering not only
in space but in time as well. When working with a differential formulation
of the conservation laws, such choices may not be as obvious.

\newpage
\section{Equations of GRMHD in Integral Form}

The numerical solutions to the equations of GRMHD follow the 
time-evolution of a fluid on a 3D spatial grid from some prescribed 
initial state at some Boyer-Lindquist time $t_i$. The numerical 
solver advances the solution in small, uniform increments $\Delta t$ 
and produces a time-ordered sequence of states for the fluid. 

The integral form of the GRMHD equations is
\begin{eqnarray}\label{conslaws}
\int_{\cal V} \nabla_\mu \,(\rho\,U^\mu) \,d^4\Omega & = 0 = & 
\int_{\partial {\cal V}} (\rho\,U^\mu)\,d^3\Sigma_\mu ,\\
\nonumber \int_{\cal V} \nabla_\mu \,T^{\mu \nu}  \,d^4\Omega & = 0 = &
\int_{\partial {\cal V}} T^{\mu \nu}\,d^3\Sigma_\mu , \\
\nonumber \int_{\cal V} \nabla_\mu \,{}^{*}F^{\mu \nu} \,d^4\Omega & = 0 = & 
\int_{\partial {\cal V}} {}^{*}F^{\mu \nu}\,d^3\Sigma_\mu  .
\end{eqnarray}
As in the previous sections, these equations will be discretized by 
taking the volume ${\cal V}$ to represent a zone on the computational
grid.

The previous section showed how an implicit scheme could be obtained from Stokes Theorem. Instead of using such a scheme, 
the GRMHD code approximates the temporal integrals by building an
approximate, upwinded 
solution at an 
intermediate time level; this approach gives rise to an explicit scheme that 
is computationally simpler. An additional motivation is that the conservation 
laws involve a mix of scalar and vectorial code variables which must be properly
centred. For example, to achieve proper centring in time and space in the case of the equation of continuity, the product of density $\rho$ (zone-centred) and 
four-velocity $U^\mu$ (face-centred) is evaluated by upwinding the density, i.e. 
by interpolating density to the appropriate zone face and advancing it in time 
to the intermediate time level, $n + 1/2$.  This estimate, at the half-step, is
then used in the continuity equation. Denote the upwinded 
quantity using an overbar; for instance, radially upwinded density would be 
shown as $\overline{\rho}^{\,r}\,U^r$. 

The details of the upwinding technique are not relevant to the present 
discussion, as they simply represent an explicit means of achieving 
proper time-centring of the integrals over the time coordinate in
(\ref{conslaws}).  A description of the
technique as it applies to the GRMHD code can be found in DH and in HSW.

\newpage
\subsection{Continuity}

The integral form of the continuity equation is
\begin{equation} 
\int_{\partial {\cal V}} (\rho\,U^\mu)\,d^3\Sigma_\mu = 0  .
\end{equation}
Using the geometry of volume ${\cal V}$ introduced earlier, the equation reads,
\begin{eqnarray} \label{continuity}
0 & = & \int_{r_0+dr} (\rho\,U^r)\,\sqrt{-g}\,dt\,d\theta\,d\phi -
       \int_{r_0} (\rho\,U^r)\,\sqrt{-g}\,dt\,d\theta\,d\phi \\
\nonumber
 & + & \int_{\theta_0+d\theta} (\rho\,U^\theta)\,\sqrt{-g}\,dt\,dr\,d\phi -
       \int_{\theta_0} (\rho\,U^\theta)\,\sqrt{-g}\,dt\,dr\,d\phi \\
\nonumber
 & + & \int_{\phi_0+d\phi} (\rho\,U^\phi)\,\sqrt{-g}\,dt\,dr\,d\theta -
       \int_{\phi_0} (\rho\,U^\phi)\,\sqrt{-g}\,dt\,dr\,d\theta \\
\nonumber
 & + & \int_{t_0+dt} (\rho\,U^t)\,\sqrt{-g}\,dr\,d\theta\,d\phi  -
       \int_{t_0} (\rho\,U^t)\,\sqrt{-g}\,dr\,d\theta\,d\phi.
\end{eqnarray}
Using the upwinding notation we have, for example,
\begin{equation} 
\int_{r_0} (\rho\,U^r)\,\sqrt{-g}\,dt\,d\theta\,d\phi \approx
\left\{\overline{\rho}^{\,r}\,U^r
\,\sqrt{-g}\,\Delta t\,\Delta \theta\,\Delta \phi \right\}_{r_0} .
\end{equation}
Using the definitions of transport velocity, $U^\mu=V^\mu\,W/\alpha$,
and auxiliary density, $D=\rho\,W$, this result is rewritten
\begin{equation} 
\int_{r_0} (\rho\,U^r)\,\sqrt{-g}\,dt\,d\theta\,d\phi \approx
\left\{\overline{D}^{\,r}\,V^r
\,\sqrt{\gamma}\,\Delta t\,\Delta \theta\,\Delta \phi \right\}_{r_0}.
\end{equation}
The integrals through the hypersurfaces of constant $t$ are approximated as \begin{equation} \label{continuity.3}
\int_{t_0+dt} (\rho\,U^t)\,\sqrt{-g}\,dr\,d\theta\,d\phi  -
       \int_{t_0} (\rho\,U^t)\,\sqrt{-g}\,dr\,d\theta\,d\phi \approx 
\left[D^{(n+1)} - D^{(n)}\right]
\,\sqrt{\gamma}\,\Delta r\,\Delta \theta\,\Delta \phi  
\end{equation}

Combining these results and simplifying yields
\begin{eqnarray} \label{continuity.4}
{D^{(n+1)} - D^{(n)} \over \Delta t} &=& -{1 \over \sqrt{\gamma}}\left[
{ \left\{\sqrt{\gamma}\,\overline{D}^{\,r}\,V^r \right\}_{r_0+\Delta\,r} - 
  \left\{\sqrt{\gamma}\,\overline{D}^{\,r}\,V^r \right\}_{r_0} \over \Delta\,r}\right.\\
\nonumber  & + & 
{ \left\{\sqrt{\gamma}\,\overline{D}^{\,\theta}\,V^\theta \right\}_{\theta_0+\Delta\,\theta} - 
  \left\{\sqrt{\gamma}\,\overline{D}^{\, \theta}\,V^\theta \right\}_{\theta_0} \over \Delta\, \theta}\\
\nonumber  & + & \left.
{ \left\{\sqrt{\gamma}\,\overline{D}^{\,\phi}\,V^\phi \right\}_{\phi_0+\Delta\,\phi} - 
  \left\{\sqrt{\gamma}\,\overline{D}^{\,\phi}\,V^\phi \right\}_{\phi_0} \over \Delta\,\phi}\right]
\end{eqnarray}
which is the discretized, upwinded form of equation (24) in DH as 
implemented in the GRMHD code, recovered very simply here by the integral 
formulation of the equation of continuity. Note how this approach ``breaks'' 
the implicit coupling at the unknown time level on 
the left- and right-hand sides shown in (\ref{flux.7}), and achieves proper 
centering through an explicit method. 

\newpage
\subsection{Induction}

The integral form of the induction equation is
\begin{equation} 
\int_{\partial {\cal V}} {}^{*}F^{\mu \nu}\,d^3\Sigma_\mu = 0.
\end{equation}
Using the geometry of volume ${\cal V}$ introduced earlier, the four
components of the induction
equation are,
\begin{eqnarray} \label{induction}
0=\int_{\partial {\cal V}} {}^{*}F^{\mu \nu}\,d^3\Sigma_\mu 
 & = & \int_{r_0+dr} {}^{*}F^{r \nu}\,\sqrt{-g}\,dt\,d\theta\,d\phi -
       \int_{r_0} {}^{*}F^{r \nu}\,\sqrt{-g}\,dt\,d\theta\,d\phi \\
\nonumber
 & + & \int_{\theta_0+d\theta} {}^{*}F^{\theta \nu}\,\sqrt{-g}\,dt\,dr\,d\phi -
       \int_{\theta_0} {}^{*}F^{\theta \nu}\,\sqrt{-g}\,dt\,dr\,d\phi \\
\nonumber
 & + & \int_{\phi_0+d\phi} {}^{*}F^{\phi \nu}\,\sqrt{-g}\,dt\,dr\,d\theta -
       \int_{\phi_0} {}^{*}F^{\phi \nu}\,\sqrt{-g}\,dt\,dr\,d\theta \\
\nonumber
 & + & \int_{t_0+dt} {}^{*}F^{t \nu}\,\sqrt{-g}\,dr\,d\theta\,d\phi  -
       \int_{t_0} {}^{*}F^{t \nu}\,\sqrt{-g}\,dr\,d\theta\,d\phi.
\end{eqnarray}

Though the previous examples did not deal with tensor quantities, the
results are readily extended.
Taking the time component, $\nu = t$, the integrals over the surfaces
of constant $t$ are identically zero (antisymmetry of $F^{\mu \nu}$), 
leaving
\begin{eqnarray} \label{induction.1}
0=\int_{\partial {\cal V}} {}^{*}F^{\mu t}\,d^3\Sigma_\mu 
 & = & \int_{r_0+dr} {}^{*}F^{r t}\,\sqrt{-g}\,dt\,d\theta\,d\phi -
       \int_{r_0} {}^{*}F^{r t}\,\sqrt{-g}\,dt\,d\theta\,d\phi \\
\nonumber
 & + & \int_{\theta_0+d\theta} {}^{*}F^{\theta t}\,\sqrt{-g}\,dt\,dr\,d\phi -
       \int_{\theta_0} {}^{*}F^{\theta t}\,\sqrt{-g}\,dt\,dr\,d\phi \\
\nonumber
 & + & \int_{\phi_0+d\phi} {}^{*}F^{\phi t}\,\sqrt{-g}\,dt\,dr\,d\theta -
       \int_{\phi_0} {}^{*}F^{\phi t}\,\sqrt{-g}\,dt\,dr\,d\theta .
\end{eqnarray}
Using (\ref{dual.2}) and substituting
the expressions for the CT variables, (\ref{ctvars}) and (\ref{maxwell.2c}),
back into (\ref{induction.1}) yields:
\begin{eqnarray} \label{induction.2}
0& = & \int_{r_0+dr} {\cal B}^r\,dt\,d\theta\,d\phi -
       \int_{r_0} {\cal B}^r\,dt\,d\theta\,d\phi \\
\nonumber
 & + & \int_{\theta_0+d\theta} {\cal B}^\theta\,dt\,dr\,d\phi -
       \int_{\theta_0} {\cal B}^\theta\,dt\,dr\,d\phi \\
\nonumber
 & + & \int_{\phi_0+d\phi} {\cal B}^\phi\,dt\,dr\,d\theta -
       \int_{\phi_0} {\cal B}^\phi\,dt\,dr\,d\theta .
\end{eqnarray}
To discretize this result, note that the vanishing of the
integrals over hypersurfaces of constant $t$ indicates that this
relation must be satisfied {\it at all times}; this component of the
induction equation is a constraint on the CT magnetic field. 
Equation (\ref{induction.2}) simplifies to:
\begin{equation} \label{induction.2a}
0 \approx {\left\{ {\cal B}^r\right\}_{r_0+\Delta\,r}-
           \left\{ {\cal B}^r\right\}_{r_0} \over \Delta r}+ 
          {\left\{ {\cal B}^\theta\right\}_{\theta_0+\Delta\,\theta}-
           \left\{ {\cal B}^\theta\right\}_{\theta_0} \over \Delta \theta}+
          {\left\{ {\cal B}^\phi\right\}_{\phi_0+\Delta\,\phi}-
           \left\{ {\cal B}^\phi\right\}_{\phi_0} \over \Delta \phi},
\end{equation}
which is the familiar ``div B'' constraint on the magnetic field, and 
equivalent to equation (47) of DH.

The spatial components of the induction equation describe the time-evolution of
the CT magnetic field. To see how this comes about, take the radial component, 
$\nu = r$, of the induction equation.
The integrals over the surfaces
of constant $r$ are identically zero, leaving
\begin{eqnarray} \label{induction.3}
0 & = & \int_{\theta_0+d\theta} {}^{*}F^{\theta r}\,\sqrt{-g}\,dt\,dr\,d\phi -
       \int_{\theta_0} {}^{*}F^{\theta r}\,\sqrt{-g}\,dt\,dr\,d\phi \\
\nonumber
 & + & \int_{\phi_0+d\phi} {}^{*}F^{\phi r}\,\sqrt{-g}\,dt\,dr\,d\theta -
       \int_{\phi_0} {}^{*}F^{\phi r}\,\sqrt{-g}\,dt\,dr\,d\theta \\
\nonumber
 & + & \int_{t_0+dt} {}^{*}F^{t r}\,\sqrt{-g}\,dr\,d\theta\,d\phi  -
       \int_{t_0} {}^{*}F^{t r}\,\sqrt{-g}\,dr\,d\theta\,d\phi.
\end{eqnarray}
Using (\ref{dual.2}), (\ref{ctvars}), and 
(\ref{maxwell.2c}), equation (\ref{induction.3}) can be discretized and 
rearranged to read
\begin{equation} 
0 \approx { {{\cal B}^r}^{(n+1)} - {{\cal B}^r}^{(n)} \over \Delta t} +
          {\left\{ \overline{\cal E}^\phi\right\}_{\theta_0+\Delta\,\theta}-
           \left\{ \overline{\cal E}^\phi\right\}_{\theta_0} \over \Delta \theta}-
          {\left\{ \overline{\cal E}^\theta\right\}_{\phi_0+\Delta\,\phi}-
           \left\{ \overline{\cal E}^\theta\right\}_{\phi_0} \over \Delta \phi}
\end{equation}
Similar rearrangements can be made for the $\theta$ and $\phi$ components of
(\ref{induction})
\begin{eqnarray} 
0&\approx&{{{\cal B}^\theta}^{(n+1)} - {{\cal B}^\theta}^{(n)} \over \Delta t} -
           {\left\{ \overline{\cal E}^\phi\right\}_{r_0+\Delta\,r}-
           \left\{ \overline{\cal E}^\phi\right\}_{r_0} \over \Delta r}+
          {\left\{ \overline{\cal E}^r\right\}_{\phi_0+\Delta\,\phi}-
           \left\{ \overline{\cal E}^r\right\}_{\phi_0} \over \Delta \phi}\\
0 & \approx &{ {{\cal B}^\phi}^{(n+1)} - {{\cal B}^\phi}^{(n)} \over \Delta t} +
           {\left\{ \overline{\cal E}^\theta\right\}_{r_0+\Delta\,r}-
           \left\{ \overline{\cal E}^\theta\right\}_{r_0} \over \Delta r}-
          {\left\{ \overline{\cal E}^r\right\}_{\theta_0+\Delta\,\theta}-
           \left\{ \overline{\cal E}^r\right\}_{\theta_0} \over \Delta \theta} 
\end{eqnarray}

These expressions are formally equivalent to the CT equations of DH (cf. equation (34) in DH). However,
as with the continuity equation, the issue of proper time-centering, this
time of the EMFs, is important. This is flagged in the above expressions by using an overbar, $\overline{\cal E}$, on the EMFs. 

Evans \& Hawley (1988) discussed the issue of time-centering of the EMFs, noting that there seemed to be some flexibility to the prescription of the EMFs. DH provided a derivation of a numerical formulation similar to the MOCCT approach of Hawley \& Stone (1995) to provide time-centred expressions which were stable under a variety of numerical tests. Perhaps 
the main advantage of revisiting CT from the point of view of Stokes Theorem
is that this formulation makes it more obvious that proper centering, both in
space and time, is indeed required. A prescription for EMFs that is compatible 
with
the integral formulation has the EMFs edge-centred on the spatial grid
and centred on the half-step on the temporal grid. The approximation using
incompressible MHD documented in DH is but one possible realization of this.
Of course, implicit formulations of (\ref{induction}) are also possible, but are 
likely to be difficult to implement.

\subsection{Energy-Momentum Conservation}

The integral form of the energy-momentum conservation equations,
\begin{equation} 
\int_{\partial {\cal V}} T^{\mu \nu}\,d^3\Sigma_\mu = 0  
\end{equation}
Using the geometry of volume ${\cal V}$ introduced earlier, the four
components of the energy-momentum equation are,
\begin{eqnarray} \label{tmn}
0=\int_{\partial {\cal V}} T^{\mu \nu}\,d^3\Sigma_\mu 
 & = & \int_{r_0+dr} T^{r \nu}\,\sqrt{-g}\,dt\,d\theta\,d\phi -
       \int_{r_0} T^{r \nu}\,\sqrt{-g}\,dt\,d\theta\,d\phi \\
\nonumber
 & + & \int_{\theta_0+d\theta} T^{\theta \nu}\,\sqrt{-g}\,dt\,dr\,d\phi -
       \int_{\theta_0} T^{\theta \nu}\,\sqrt{-g}\,dt\,dr\,d\phi \\
\nonumber
 & + & \int_{\phi_0+d\phi} T^{\phi \nu}\,\sqrt{-g}\,dt\,dr\,d\theta -
       \int_{\phi_0} T^{\phi \nu}\,\sqrt{-g}\,dt\,dr\,d\theta \\
\nonumber
 & + & \int_{t_0+dt} T^{t \nu}\,\sqrt{-g}\,dr\,d\theta\,d\phi  -
       \int_{t_0} T^{t \nu}\,\sqrt{-g}\,dr\,d\theta\,d\phi.
\end{eqnarray}

The discretization of this set of equations proceeds as described above.
The only issue is the relation between the primitive code variables
and $T^{\mu \nu}$, and the extraction of the primitive variables. 

In the GRMHD code, the energy-momentum conservation law is treated differently, and the resulting equations provide a so-called non-conservative treatment of
energy. The conservation law $\nabla_\mu\,T^{\mu \nu}=0$ is applied to a
fluid with four-velocity $U^\mu$. A very convenient reformulation of the
conservation law arises when one ``reorients'' the equations into a component parallel to the local four-velocity and components orthogonal to it (MTW). 
Projection onto the four-velocity yields a scalar relation,
\begin{equation}\label{proj}
U_\nu\,\nabla_\mu\,T^{\mu \nu}=0,
\end{equation}
which, as shown in HSW and DH, leads to the equation of internal energy 
conservation,
\begin{equation} \label{enfinal}
 \partial_{t}\left(E\right)+{1 \over \sqrt{\gamma}}\,
\partial_{i}\left(\sqrt{\gamma}\,E\,V^i\right)
+ P\,\partial_{t}\left(W\right) + 
{P\over\sqrt{\gamma}}\,\partial_{i}\left(\sqrt{\gamma}\,W\,V^i\right)
= 0 .
\end{equation}
To obtain the components of the conservation law orthogonal to the four-velocity, use the projection tensor $P_{\mu \nu} = g_{\mu \nu}+U_\mu\,U_\nu$,
\begin{equation}\label{mom}
P_{\alpha \nu}\,\nabla_\mu\,T^{\mu \nu}=
g_{\alpha \nu}\,\nabla_\mu\,T^{\mu \nu}+U_\alpha\,U_\nu\,\nabla_\mu\,T^{\mu \nu}
= \nabla_\mu\,{T^\mu}_\alpha = 0,
\end{equation}
where the final result was obtained by commuting the metric with the covariant 
derivative and also using (\ref{proj}). As shown in DH, the three spatial
components of this equation are
\begin{eqnarray}\label{mom.3}
0 & = &\partial_t\left(S_j-\alpha\,b_j\,b^t\right)\\
\nonumber & + &  {1 \over \sqrt{\gamma}}\,
  \partial_i\,\sqrt{\gamma}\,\left(S_j\,V^i-\alpha\,b_j\,b^i\right)+
  {1 \over 2}\,\left({S_\epsilon\,S_\mu \over S^t}-
  \alpha\,b_\mu\,b_\epsilon\right)\,
  \partial_j\,g^{\mu\,\epsilon}+
  \alpha\,\partial_j\left(P+{{\|b\|}^2 \over 2}\right).
\end{eqnarray}
The time-component is not used; instead, the GRMHD code uses the normalization condition
\begin{equation}\label{momnorm}
g^{\mu \nu}\,S_\mu\,S_\nu = -{\left[(\rho\,h+{\|b\|}^2)\,W\right]}^2 
\end{equation}
to close the set of equations.

\section{Summary and Discussion}

This paper has outlined the development of discretized equations of ideal
General Relativistic Magnetohydrodynamics (GRMHD) from the integral form
of the conservation laws. Stokes Theorem transforms the conservation
laws of GRMHD into simple flux integrals through the boundary of a closed 
four-volume, and many possible discretizations are possible for these integrals. 
The particular choices used in the GRMHD code for the equation of continuity and the induction equation represent but one implementation,
arguably the simplest and most straightforward of these. Many other possible
implementations are implied by equations (\ref{continuity}) and
(\ref{induction}). The GRMHD code uses an alternate
discretization of the energy-momentum conservation equation; the
integral formulation discussed here, (\ref{tmn}), offers another possible discretization in keeping with the treatment of the continuity and
induction equations.
A note of caution is in order: the integral relations derived here, dealing as they do with coupled sets of non-linear equations, may not yield stable 
numerical schemes. Since only limited stability analysis is possible in 
non-linear differential equations, theoretical work must be
complemented by careful development and testing to validate any particular 
numerical treatment.

In addition, this discussion has implicitly assumed smooth flows; the
issue of shock capturing has not been touched upon. The GRMHD code uses
a relativistic treatment that is based on the original Von Neumann \&
Richtmyer (1950) artificial viscosity treatment, a relativistic
extension of this technique by May \& White (1967), and HSW. This remains the 
state of the art as of this writing.

As a final note, the zone-based, flux-conserving formulation discussed here may also be of 
benefit in treating boundary conditions. The boundary zones, or ghost zones, 
of the computational grid are used to impose physical boundary conditions on
the computational volume and also to provide approximate data used
by the interpolation routines to provide correct time-centering of
zones on the computational grid adjacent to the physical boundaries.
The main interest in the integral formulation of boundary
data lies at the inner radial boundary, where frame dragging effects are 
important, and also at the difficult ``corner zones'' which sit just outside
the pole caps of the black hole. Definitive statements regarding the
flux of electromagnetic energy near the event horizon hinge on reliable
boundary treatments, and it is hoped that a more sophisticated
approach, based on the integral formulation discussed here, may help. 
This remains an open area of research.

\noindent{\bf Acknowledgements:}

\noindent The GRMHD code was developed by the author while at the University of 
Virginia, under NSF grant AST-0070979 and NASA grant NAG5-9266 (J.F. Hawley and
S. Balbus, Principal Investigators). A small portion of this
work was also carried out while the author was at the University of Calgary,
partially supported by the Natural Sciences and Engineering Research
Council of Canada (NSERC), as well as the Alberta Ingenuity Fund (AIF).

%%
%%%%%%%%%%%%%%%%%%%%%%%%%%%%%%%%%%%%%%%%%%%%%%%%%%%%%%%%%%%%%%%%%%%%%%%%%%
%%
%% REFERENCES  
%%


\begin{thebibliography}{99}
\bibitem{A89} Anile, A.~M. 1989, 
 Relativistic Fluids and Magnetofluids (Cambridge: Cambridge University Press)
\bibitem {C96} Clarke, D.~A. 1996, ApJ, 457, 291
\bibitem{DH} De Villiers, J.~P. \& 
 Hawley, J.~F. 2003, ApJ, 589, 458 (DH)
\bibitem{DHK} De Villiers, J.~P., 
 Hawley, J.~F., \& Krolik, J.~H. 2003, ApJ, 599, 1238 (DHK)
\bibitem{D:05} De Villiers, 
J.~P., Hawley, J.~F., Krolik, J.~H., \& Hirose, S. 2005, ApJ, 620, 878
\bibitem{D06} De Villiers, J.~P. 2006, astro-ph/0605744
\bibitem{EH:88} Evans, C.~R. \& Hawley, J.~F.
 1988, ApJ,  332, 659 (EH88)
\bibitem{HSW1:84} 
 Hawley, J.~F., Smarr, L.~L., \& Wilson, J.~R., 1984, ApJ, 277, 296 (HSW)
\bibitem{HS95} 
 Hawley, J.~F. \& Stone, J.~M. 1995, Comput. Phys. Comm., 89, 127
\bibitem{PaperII} 
 Hirose, S., Krolik, J.~H., De Villiers, J.~P., \& Hawley, J.~F. 2004, ApJ, 
 606, 1083
\bibitem{L67} Lichnerowicz, A. 1967, 
 Relativistic Magnetohydrodynamics (New York: Benjamin)
\bibitem{MTW} Misner, C.~W., Thorne,
 K.~S., \& Wheeler, J.~ A, 1973, Gravitation (San Francisco: W.H. Freeman)
\bibitem{MW} May, M. M., \& White, R. H. 1967. In {\it Methods of Computational
Physics}, Vol, 7, ed. B. Adler, S. Fernbach, and M. Rolenberg (New York: Academic), p. 219.
\bibitem{pt74}Page, D. N., \& Thorne, K. S.
 1974, ApJ, 191, 499
\bibitem{VNR:50} 
 Von Neumann, J., \& Richtmyer, R.D., 1950, Journal of Applied Physics, 
 21, 232
\end{thebibliography}
\end{document}